\documentclass[a4paper,fleqn]{cas-dc}

\usepackage[numbers]{natbib}

\def\tsc#1{\csdef{#1}{\textsc{\lowercase{#1}}\xspace}}
\tsc{WGM}
\tsc{QE}

\newcommand{\Palpha}{$P^{\alpha}_{\rm{tail}}$}

\begin{document}
\let\WriteBookmarks\relax
\def\floatpagepagefraction{1}
\def\textpagefraction{.001}

\shorttitle{Cherenkov telescopes
enhanced by a ground array observatory}    

\shortauthors{C. Alispach et~al.}  

\title[mode = title]{Prospects for the detection of gamma rays using Cherenkov telescopes enhanced by a ground array observatory}

%

\affiliation[1]{organization={D\'epartement de Physique Nucl\'eaire, Facult\'e de Sciences, Universit\'e de Gen\`eve}, addressline={24 Quai Ernest Ansermet}, city={Gen\`eve}, postcode={1205}, country={Switzerland}}

\affiliation[2]{organization={FZU - Institute of Physics of the Czech Academy of Sciences}, addressline={Na Slovance 1999/2}, city={Prague}, postcode={18200}, country={Czech Republic}}

\affiliation[3]{organization={Pidstryhach Institute for Applied Problems of Mechanics and Mathematics, National Academy of Sciences of Ukraine}, addressline={3-b Naukova St.}, city={Lviv}, postcode={79060}, country={Ukraine}}

\affiliation[4]{organization={Nicolaus Copernicus Astronomical Center, Polish Academy of Sciences}, addressline={ul. Bartycka 18}, city={Warsaw}, postcode={00-716}, country={Poland}}

\affiliation[5]{organization={Institute of Nuclear Physics, Polish Academy of Sciences}, city={Krakow}, postcode={31-342}, country={Poland}}

\affiliation[6]{organization={Astronomical Observatory, University of Warsaw}, addressline={Al. Ujazdowskie 4}, city={Warsaw}, postcode={00-478}, country={Poland}}

\affiliation[7]{organization={Palack\'y University Olomouc, Faculty of Science}, addressline={17. listopadu 50}, city={Olomouc}, postcode={77900}, country={Czech Republic}}

\affiliation[8]{organization={Deutsches Elektronen-Synchrotron (DESY)}, addressline={Platanenallee 6}, city={Zeuthen}, postcode={15738}, country={Germany}}

\affiliation[9]{organization={Faculty of Physics, University of Bia{\l}ystok}, addressline={ul. K. Cio{\l}kowskiego 1L}, city={Bia{\l}ystok}, postcode={15-245}, country={Poland}}

\affiliation[10]{organization={Astronomical Institute of the Czech Academy of Sciences}, addressline={Fri\v{c}ova~298}, city={Ond\v{r}ejov}, postcode={25165}, country={Czech Republic}}

\affiliation[11]{organization={Astronomical Institute of the Czech Academy of Sciences}, addressline={Bo\v{c}n\'i~II 1401}, city={Prague}, postcode={14100}, country={Czech Republic}}

\affiliation[12]{organization={D\'epartement d'Astronomie, Facult\'e de Science, Universit\'e de Gen\`eve}, addressline={Chemin d'Ecogia 16}, city={Versoix}, postcode={1290}, country={Switzerland}}

\affiliation[13]{organization={ETH Zurich, Institute for Particle Physics and Astrophysics}, addressline={Otto-Stern-Weg 5}, city={Zurich}, postcode={8093}, country={Switzerland}}

\affiliation[14]{organization={Institute of Particle and Nuclear Physics, Faculty of Mathematics and Physics, Charles University}, addressline={V Hole\v sovi\v ck\' ach 2}, city={Prague}, postcode={18000}, country={Czech~Republic}}

\affiliation[15]{organization={Astronomical Observatory, Jagiellonian University}, addressline={ul. Orla 171}, city={Krakow}, postcode={30-244}, country={Poland}}

\affiliation[16]{organization={Physics Department, Instituto Superior T\'ecnico, Universidade de Lisboa and LIP - Laboratório de Instrumentação e Física Experimental de Partículas}, addressline={Av. Prof. Gama Pinto 2}, city={Lisboa}, postcode={1649-003}, country={Portugal}}

\author[1]{C. Alispach}
\author[2]{A. Araudo}
\author[2]{A. Bakalov\'{a}}
\author[1]{M. Balbo}
\author[3]{V. Beshley}
\author[2]{J. Bla\v{z}ek}
\author[4]{J. Borkowski}
\author[6]{T. Bulik}
\author[1]{F. Cadoux}
\author[5]{S. Casanova}
\author[2]{A. Christov}
\author[2]{J. Chudoba}
\author[7]{L. Chytka}

\author[2]{P. \v{C}echvala}[orcid=0009-0009-2107-1848]
\cormark[1]
\ead{cechvala@fzu.cz}

\author[2]{P. D\v{e}dic}
\author[1]{Y. Favre}
\author[8]{M. Garczarczyk}
\author[9]{L. Gibaud}
\author[5]{T. Gieras}
\author[9]{E. G{\l}owacki}
\author[7]{P. Hamal}
\author[1]{M. Heller}
\author[7]{M. Hrabovsk\'y}
\author[2]{P. Jane\v{c}ek}
\author[10]{M. Jel\'inek}
\author[7]{V. J\'ilek}
\author[2]{J. Jury\v{s}ek}
\author[11]{V. Karas}
\author[1]{B. Lacave}
\author[12]{E. Lyard}
\author[2]{D. Mand\'at}
\author[5]{W. Marek}
\author[7]{S. Michal}
\author[5]{J. Micha{\l}owski}
\author[9]{M. Miro\'n}
\author[4]{R. Moderski}
\author[1]{T. Montaruli}
\author[4]{A. Muraczewski}
\author[2]{S. R.~Muthyala}
\author[2]{A.~L. Müller}
\author[1]{A. Nagai}
\author[5]{K. Nalewajski}
\author[13]{D. Neise}
\author[5]{J. Niemiec}
\author[9]{M. Niko{\l}ajuk}

\author[2,14]{V. Novotn\'y}[orcid=0000-0002-4319-4541]
\cormark[1]
\ead{vladimir.novotny@matfyz.cuni.cz}

\author[15]{M. Ostrowski}
\author[2]{M. Palatka}
\author[2]{M. Pech}
\author[2]{M. Prouza}
\author[2]{P. Schov\'{a}nek}
\author[2]{T. Schulz}
\author[12]{V. Sliusar}
\author[10]{J. Srba}
\author[15]{{\L}. Stawarz}
\author[8]{R. Sternberger}
\author[1]{M. Stodulska}
\author[5]{J. \'{S}wierblewski}
\author[5]{P. \'{S}wierk}
\author[10]{J. \v{S}trobl}
\author[2]{T. Tavernier}
\author[2]{P. Tr\'avn\'i\v{c}ek}
\author[1]{I. Troyano~Pujadas}
\author[2]{J. V\'icha}
\author[12]{R. Walter}
\author[15]{K. Zi{\c e}tara}
\author{\bf{(SST-1M~Collaboration)}}

\author[16]{R. Conceição}
\author[16]{L. Gibilisco}
\author[16]{M. Pimenta}
\author[16]{B. Tomé}

\cortext[1]{Corresponding author}

\begin{abstract}
We study through detailed simulated data and their optimized analysis the expected performance of the Single-Mirror Small-Size imaging atmospheric Cherenkov Telescopes (SST-1M) potentially located inside a high-altitude array of Water-Cherenkov Detectors (WCDs) inspired by the current foreseen design of the Southern Wide-field Gamma-ray Observatory (SWGO).
For such a hybrid setup, we show an improvement in the flux sensitivity above 10\,TeV by about 60\% for monocular and 30\% for stereoscopic SST-1M observation, due to the improved gamma/hadron separation when additional parameters from the WCD array are used.
We also discuss further benefits of the hybrid gamma observatory concept and its technical challenges.
\end{abstract}


\maketitle

\section{Introduction}
\label{intro}

Observations of very-high-energy (VHE) gamma rays provide crucial insights into the most energetic astrophysical accelerators.
Ground-based VHE gamma-ray detection relies on particle-detector arrays or imaging atmospheric Cherenkov telescopes (IACTs).
However, each technique comes with its limitations. While particle-detector arrays provide a wide field of view and a high duty cycle with limited angular resolution, IACTs have excellent 
angular and energy resolution with a limited field of view and duty cycle. 
We study a possible hybrid detection approach that combines a particle-detector array with IACTs, offering, in a joint mode, an improved sensitivity and in a disjoint mode the possibility to operate for longer duty cycles and larger FoV than what IACTs can do. The potential of such a hybrid detection configuration has already been explored by the LHAASO Collaboration which is deploying the LACT IACT array close to the LHAASO pond of water and particle-detector array~\citep{LHAASOhybrid}.

In this paper, we present a simulation study of a hybrid-detection concept combining two IACTs---modeled as Single-Mirror Small-Size Telescopes (SST-1Ms)~\citep{SST-1M, Alispach:2025mbz} utilizing a silicon-photomultiplier camera~\citep{2017EPJC...77...47H}---with a simplified particle-detector array inspired by the design concept of the Southern Wide-field Gamma-ray Observatory (SWGO)~\citep{SWGO,SWGO:2025taj}.
Preliminary results of this study were presented in Ref.~\citep{HybridICRC}.
Although the benefit is mutual for a particle-detector array and an array of IACTs, in this paper, we focus on determining the improvement of the sensitivity of two SST-1M IACTs due to being located in the middle of particle detectors.

The paper is structured as follows: Section~\ref{sec:sim} describes the simulation setup;
Section~\ref{sec:ground} presents the observables used to enhance gamma/hadron ($\gamma$/h) separation capability; the main results concerning the sensitivity of SST-1M in monocular and stereoscopic modes, when combined with a particle-detector array, are presented in Section~\ref{sec:sensitivity}.
The general discussion of scientific benefits of such a hybrid array is given in Section~\ref{sec:hybrid}; the technical aspects of a high-altitude operation of SST-1M are briefly discussed in Section~\ref{sec:tech}. 
Section~\ref{sec:sum} summarizes the obtained findings and describes the future outlook of this work.

\section{Simulation setup}
\label{sec:sim}

In order to evaluate the benefit of the addition of an array of water-Cherenkov detectors (WCDs) to IACTs, we prepared a three-step set of detailed Monte Carlo (MC) simulation.

In the first step, the development of air showers and the consequent production of Cherenkov light is simulated with the  \texttt{CORSIKA\,v7.7402}~\citep{1998cmcc.book.....H} framework.
We employed the hadronic interaction model \texttt{UrQMD}~\citep{urqmd2} for low energies (<80\,GeV) and \texttt{QGSJet\,II-04}~\citep{qgs} was used for interactions at higher energies.
In the second step, the response of the SST-1M telescopes was simulated with \texttt{sim\_telarray\,v2021-12-25} \citep{BERNLOHR2008149}, also taking into account the light attenuation in the atmosphere.  
In the third step, the WCD signals were evaluated using the simplified framework adopted from Ref.~\citep{Conceicao:2022lkc_LCm}.
Within this framework, cylindrical WCDs are placed on a regular grid and the signal from through-going particles is calculated from a GEANT4-based parameterization of the response of individual WCDs, taking into account fluctuations of the signal.

\begin{figure}
    \centering
    \includegraphics[width=1.0\linewidth]{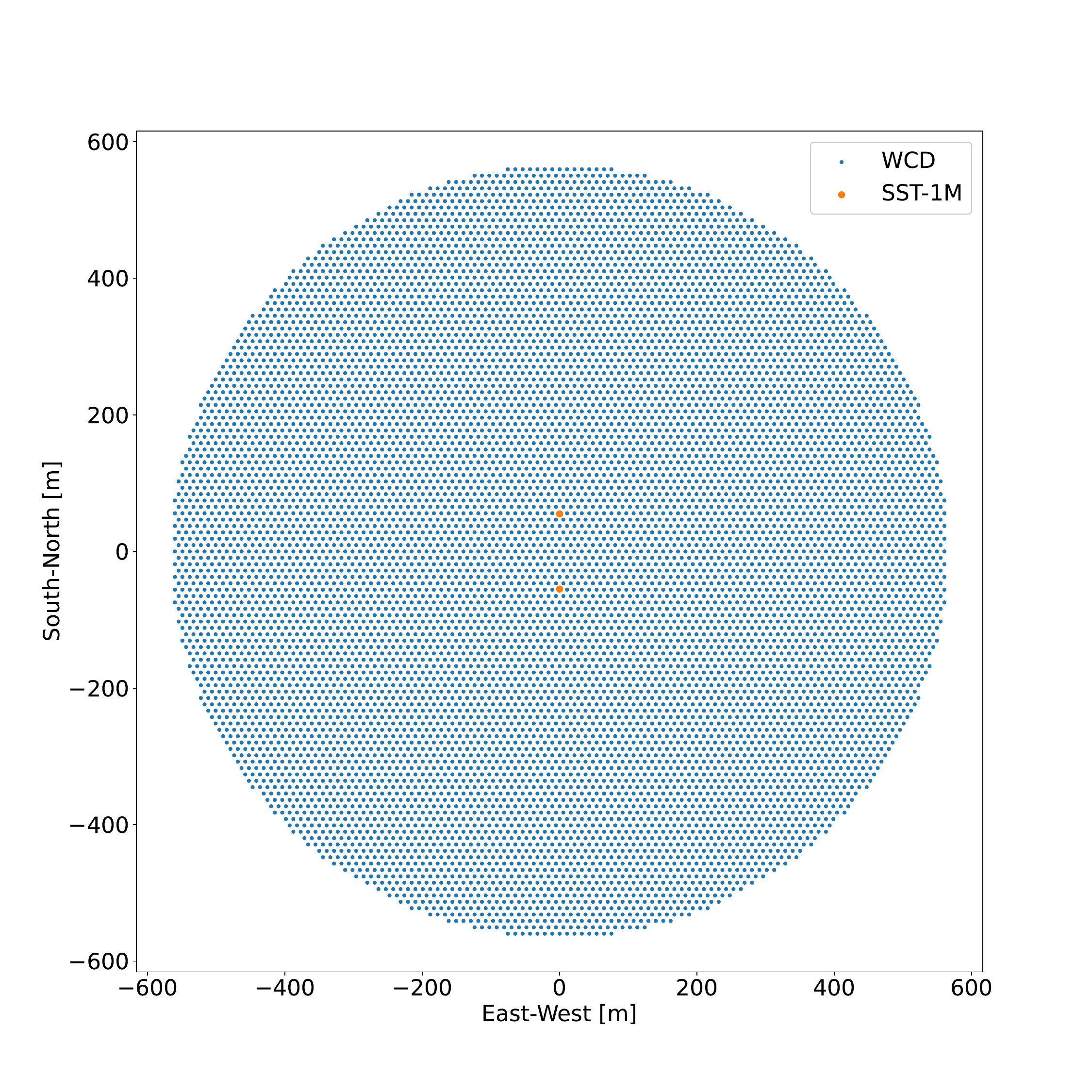}
    \caption{Sketch of the array of WCDs together with positions of 2 SST-1Ms.}
    \label{fig:array}
\end{figure}

The particle-detector array consists of 9997 WCDs, each 
with an area of 12.6 m$^2$ and equipped with 3 photomultiplier tubes.
The detectors are distributed over a total area of approximately 1 km$^2$ in a uniform triangular grid, shown in Fig.~\ref{fig:array}, resulting in a total area coverage of 12.5\%.
The very large number of WCDs, combined with the relatively high fill factor (FF)\footnote{The FF is defined as total active detector area/total instrumented area, where the FF corresponds to distances between tanks of 10.8\,m.} of 12.5\%, was deliberately chosen to minimize border effects and potential reconstruction artifacts.
By comparison, the expected SWGO array features FFs of 70\%, 4\%, and 1.7\%, decreasing with radius and defining the SWGO zones with radii of 160\,m, 400\,m and 560\,m for zone 1, 2 and 3, respectively~\citep{SWGO:2025taj}.
This changing distance between tanks is optimized in order to decrease the array cost while preserving capability to fulfill the scientific goals of the observatory.

Motivated by the future SWGO site~\citep{SWGO:2025taj}, we set the simulated altitude to 4700\,m and 
adapted in our simulation the appropriate value of the density profile to this altitude and the geomagnetic field to the Pampa la Bola site located in the Atacama Astronomical Park, Chile.
The atmosphere transmissivity set in \texttt{sim\_telarray} was calculated using MODTRAN~\citep{modtran} for the same site.

Two SST-1M telescopes were placed symmetrically around the center of the particle-detector array, 110\,m apart in the North–South direction. Since the array extends up to 565\,m from its center, each telescope is surrounded by a region of approximately 500\,m radius instrumented with $\approx$7800 WCDs. This scale is comparable to the impact-parameter range over which SST-1M telescopes contribute to the effective area at energies around 100\,TeV.
We checked that percent-level differences in sensitivities arise for other orientations of the array layout.
Both telescopes were used to obtain stereoscopic results, while
the SST-1M that is more towards South was masked during event processing to obtain monocular performance.
Due to the lack of on-site measurements of a night sky background (NSB) photon rate at Pampa La Bola, we conservatively assume in the present simulations the NSB rate of 72\,MHz corresponding to the darkest nights detected by the SST-1M at Ond\v rejov, where the telescopes currently operate.
This measurement-validated value is close to 80--92 MHz that we estimated in our previous NSB simulations for a similar desert-area site of Paranal, although at lower altitude.
Showers with a zenith angle of 20\textdegree, coming from the North, were simulated to benchmark low zenith angle observations.

The IACT simulations were processed using an open source software \texttt{sst1mpipe~v0.7.4}\footnote{\url{https://github.com/SST-1M-collaboration/sst1mpipe}}~\citep{jurysek_2025_14808846} developed for the calibration and air-shower reconstruction of SST-1M. Minor modifications to the used version of \texttt{sst1mpipe} were made in order to account for the selected site.
Random Forest (RF) methods \citep{Breiman:2001hzm}, used for event reconstruction, were trained on MC-simulated diffuse protons and gamma rays, while the response to a point-like source was calculated using point-like gammas.
The primary particles, generated according to the energy spectral shape $\frac{\rm{d}N}{\rm{d}E} \propto E^{-2}$, have energies between 200\,GeV and 631\,TeV, and 400\,GeV and 1100\,TeV for gamma rays and protons, respectively, to account for the migration of background events into the analyzed energy range.
For sensitivity calculations, the proton distribution is further reweighted to the DAMPE-measured p+He spectrum~\cite{DeMitri:2021jog}.
To cover the whole 9\textdegree~field of view of SST-1M in the case of diffuse samples, showers are simulated up to 5\textdegree~from the pointing axis of the telescopes.
Events are evenly distributed in the area up to the impact distance of 1032\,m from the array center, while the radius of the 1\,km$^2$ particle-detector array was 565\,m.
Assuming that only showers with a core inside the WCD array can be reasonably analyzed by particle detectors,
events with a core distance greater than 565\,m are not used in the WCD array-enhanced analysis.
In this case, only about $\approx$2\% of the reconstructed gamma rays are discarded, demonstrating that the array size is sufficient to adequately cover the effective area of the SST-1M telescopes in the configuration considered.
Nevertheless, we keep off-array events for the IACT-only processing.

The baseline SST-1M reconstruction uses an RF $\gamma$/h classifier to assess the {\it gammaness} parameter~\citep{Alispach:2025mbz}, which represents how closely a particular shower resembles the expectations from gamma-ray showers.
The following Hillas variables are used to train the classifier, namely the log of the intensity, width, length, width/length ratio, skewness, kurtosis, timing slope, leakage, and coordinates of the shower center of gravity in the FoV (x,y).

A highly performing variable driving the classification for gamma-hadron discrimination is Hillas’ width, which measures the spread of a shower on the camera; see Ref.~\citep{Alispach:2025mbz} where also the energy and angular reconstructions, performed in a similar way using dedicated RF regressors, are explained.
In the presented study, the energy and angular reconstructions are performed solely using SST-1M.

The stereoscopic reconstruction method uses, independently for each telescope, the RF variables listed above for the $\gamma$/h classifier, together with the distance of the impact point from the telescopes and the distance from the center of the IACT array to the shower maximum, $h_{\rm{max}}$.
The results of the two estimates are then averaged using the image intensity as a weighting factor.
More details on the implementation of stereoscopic reconstruction are given in~Ref.~\citep{Alispach:2025mbz}.

\section{Particle-detector array observables}
\label{sec:ground}

In this work, we use the information from the particle-detector array to enhance the $\gamma$/h separation capability of the SST-1M telescopes.
This enhancement stems mainly from the sensitivity of the WCD array to the muon content of air showers, which is almost absent for gamma-ray primaries but substantial for those of hadronic origin.
To study potential improvement, we first focus on adding the $\gamma$/h separation variables obtained from the WCD array to the standard SST-1M analysis. Exploitation of the complete ground-array information could yield better performance but would be the subject to further work and is discussed in Section~\ref{sec:hybrid}.
Instead, we extracted two $\gamma$/h separation variables, $LCm$ and \Palpha, using a simplified framework from Ref.~\citep{Conceicao:2022lkc_LCm}, under the assumption that the shower footprints are contained within the array.
These variables were chosen due to their great primary separation performance and easy applicability to a variety of different array configurations and particle-detector realizations considered for future WCD-based observatories.

The first variable, $LCm$ \citep{Conceicao:2022lkc_LCm}, quantifies the azimuthal fluctuations of the ground signal.
On the one hand, gamma-ray induced showers are dominated by the electromagnetic component and therefore produce a rather smooth lateral signal distribution.
On the other hand, hadronic showers exhibit stronger azimuthal fluctuations of the signal due to their more complex structure.
Thanks to this difference, quantified by $LCm$, the variable is an excellent $\gamma$/h separator.
For a detailed definition of the $LCm$ variable and its performance, see Refs.~\citep{Conceicao:2022lkc_LCm,Conceicao:2023LCm, LCmPeV}.

The second variable extracted from the WCD array is \Palpha~\citep{Conceicao:2023ybu_Ptail}, which represents the extent to which the detectors record a signal significantly above the average of all detectors at a comparable core distance, which indicates muon charge release.
This parameter has been shown to be well correlated with the total number of muons and thus provides an efficient $\gamma$/h separation variable that allows us to obtain high-purity gamma samples.

Both variables are reconstructed for shower energies greater than 10\,TeV, as suggested in Ref.~\citep{Conceicao:2022lkc_LCm}, while below 10\,TeV the $\gamma$/h separation relies only on the rest of the parameters.
However, other methods and variables that correlate with the number of muons can also be introduced at lower energies, as shown in Refs.~\citep{SWGO:2025taj, LHAASO, Conceicao:2021xgn}.
Thus, in addition to the two variables reconstructed from the particle-detector array, we also use for illustration purposes the true number of muons in the showers, $N_{\mu}$, obtained directly from CORSIKA simulations at the ground level within a 1\,km$^2$ circle around the shower core.
Although this variable is not usable in real data, it provides insights into which of the proxies more closely approach rejection efficiency.

\section{SST-1M with enhanced gamma/hadron separation}
\label{sec:sensitivity}

In this section, we quantify the expected performance improvement of SST-1M telescopes when included in a particle-detector array.
We focus on the impact of the additional WCD-array information on $\gamma$/h separation and on the resulting sensitivity of the combined system, presenting the results separately for the monocular and stereoscopic SST-1M operation modes.

The evaluation is performed by comparing the baseline SST-1M
reconstruction with the hybrid setup described in Sections~\ref{sec:sim}
and \ref{sec:ground}.
The WCD-array discrimination variables, \Palpha\ and $LCm$, were incorporated as additional features in the RF classifier trained within \texttt{sst1mpipe} to perform $\gamma$/h separation.
The trained RF model was subsequently applied to an independent test MC sample.
Because the parameterizations used to calculate $LCm$ and \Palpha\ are energy dependent, we use the reconstructed energy by a separate RF regressor rather than the true energy from the MC to calculate \Palpha\ and $LCm$, so to propagate effects related to the energy reconstruction process into these variables.

To assess the separation capability of the hybrid approach, we employ Receiver Operating Characteristic (ROC) curves and their integral metrics to enable a direct comparison with the baseline configuration.
To evaluate the final performance of the SST-1M telescopes, we apply standard event-quality selections following Ref.~\citep{Alispach:2025mbz}: a Hillas' intensity cut of >45 photoelectrons and a leakage cut of <0.7.
An energy-dependent gammaness cut is then applied to retain a fixed efficiency for gamma-ray selection of 60\% in both the monocular and stereoscopic analyses, while imposing an upper limit of 0.95 on the cut value.
Finally, to maximize the detection significance, i.e., to improve the sensitivity to a point-like source at $20^{\circ}$ zenith as much as possible, an energy-dependent cut on the distance between the reconstructed shower direction and the true source position, $\theta^2$, is independently optimized.
The optimization process resulted in a cut that further reduced the number of selected gammas.
The best performance was achieved for 50\% and 68\% gamma-ray selection efficiency for monocular and stereoscopic analysis, respectively.
This selection was applied on top of the already applied gammaness cut, resulting in the combined ratio of selected gammas of 30\% and 40.8\% for monocular and stereoscopic analysis, respectively.

Finally, we translate the changes in $\gamma$/h separation performance into sensitivity estimates for the hybrid array and discuss the conditions under which the hybrid concept yields the largest gains with respect to the baseline SST-1M reconstruction.
Because the SST-1M performance depends strongly on the telescope trigger multiplicity, we present the results for monocular and stereoscopic configurations separately. 

\subsection{Monocular performance}
\label{sec:mono}

\begin{figure}
    \centering
    \includegraphics[width=1.0\linewidth]{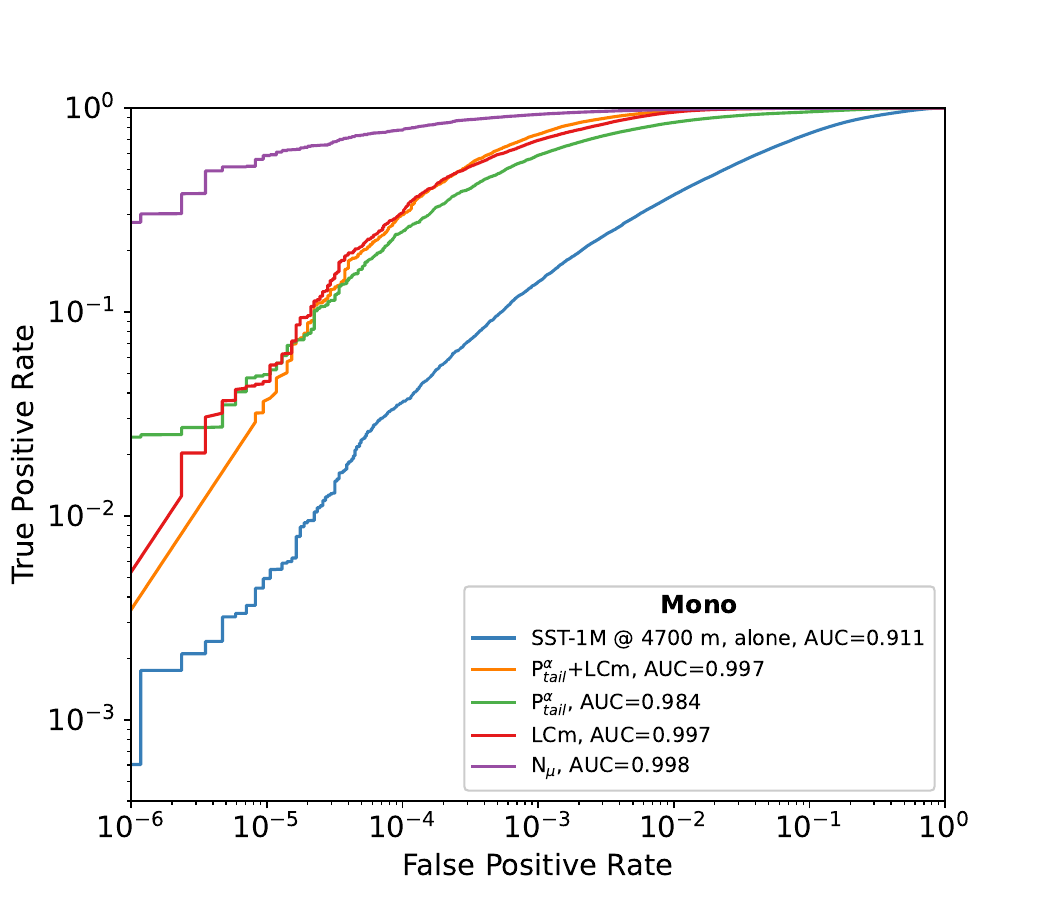}
    \caption{ROC curves in the monocular observation regime.
    Shown for a single SST-1M telescope alone (blue) and for SST-1M together with $\gamma$/h discrimination parameters obtained from the particle-detector array, including \Palpha\ (green), $LCm$ (red), both \Palpha\ and $LCm$ (orange), and true number of muons $N_{\mu}$ (purple).  
    All curves are shown for energies above 10\,TeV.}
    \label{fig:roc_mono}
\end{figure}

In this Section, we estimate the performance of a single SST-1M telescope in the middle of the particle-detector array.
Since the \Palpha\ and $LCm$ discriminators are, by design, intended to operate at high energies---typically above 10\,TeV---the ROC curves shown in Fig.~\ref{fig:roc_mono} are evaluated only for events with reconstructed energies above this energy, ensuring a fair and meaningful comparison.

\begin{figure}
    \centering
    \includegraphics[width=1.0\linewidth]{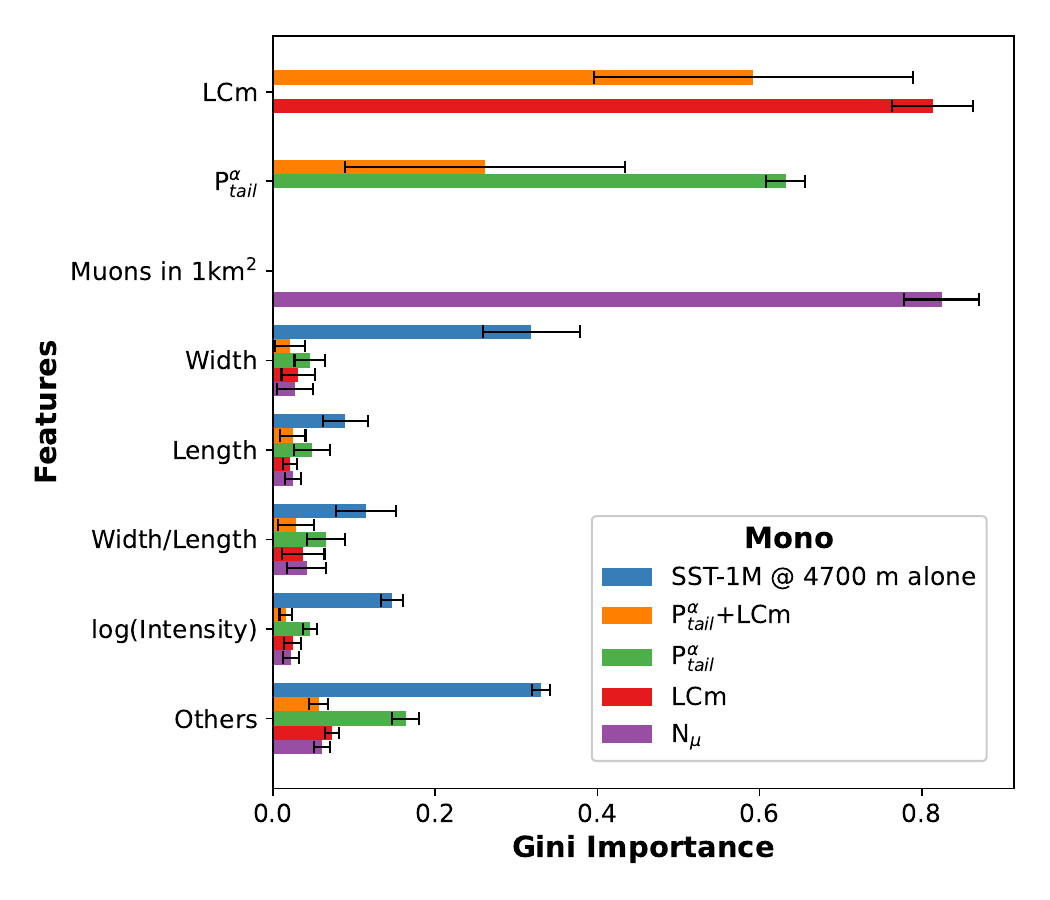}
    \caption{Gini importance of features used in the Random Forests classifiers that separate gamma rays from hadrons.
    The analyses that use only the monocular SST-1M (blue) and SST-1M telescope with additional information from the particle-detector array are shown: true $N_{\mu}$ (purple), $LCm$ (red), \Palpha\ (green), and \Palpha+$LCm$ (orange).
    Only the most important features are depicted, while the contribution of the rest of them, listed in the text, is summed and shown as Others.}
    \label{fig:gini_mono}
\end{figure}

To illustrate how much information in the gammaness comes from individual variables, we show the Gini importance \citep{Breiman:2001hzm} of individual reconstruction parameters for different $\gamma$/h classifiers in Fig.~\ref{fig:gini_mono}.
Gini importance is a metric used in RF models to measure the relevance of input features.
It does so by quantifying the contribution of each variable to reducing uncertainty during the training process.
The importance is calculated by summing up the weighted total reduction in node impurity attributed to a particular feature across all trees in the ensemble.
The weight corresponds to the probability that samples reach corresponding nodes.
Along the y-axis of Fig~\ref{fig:gini_mono}, only the most important variables are indicated, while a larger number of them, listed in Section~\ref{sec:sim}, is used and set together under `Others'.
The colored bars show the Gini importance for the SST-1M in monocular mode, and adding information from the particle detectors.

The decision relies heavily on the number of muons, $N_{\mu}$, when included, consistent with the fact that gamma-ray showers are expected to produce fewer muons than hadronic showers.
Similarly, \Palpha\ and $LCm$ also dominate the decision logic overall when used due to their correlation with $N_{\mu}$.
When both $LCm$ and \Palpha\ are present in the analysis, $LCm$ is more important than \Palpha, demonstrating its better $\gamma$/h separation capability in this particular dense layout.

\begin{figure}
    \centering
    \includegraphics[width=1.0\linewidth]{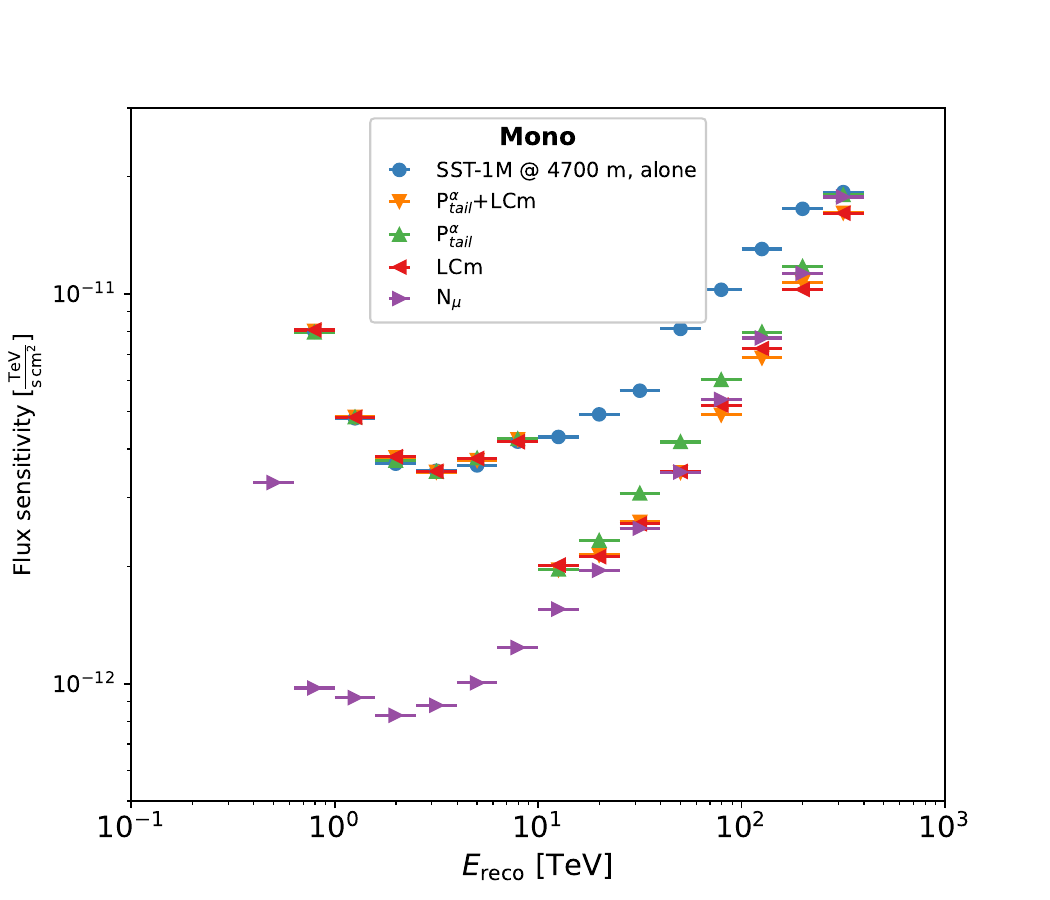}
    \caption{Flux sensitivity for a sole SST-1M telescope (blue) and one SST-1M with the additional information from the particle-detector array.
    The color coding is the same as in Fig~\ref{fig:roc_mono}.
    A point-like source at the zenith angle of 20\textdegree~observed for 50\,h is assumed.
    }
    \label{fig:sensitivity_mono}
\end{figure}

Using the RF models built above, we evaluated the sensitivity of a single SST-1M to a point-like source at 20\textdegree~zenith angle, observed for 50\,h, utilizing different $\gamma$/h separation methods.
The sensitivity is defined according to the standard rules in gamma astronomy, presented in Ref.~\citep{CTAConsortium:2012yjm}, as a flux for which the detection with 5$\sigma$ statistical significance should be achieved, at least ten excess events surviving cuts are demanded, and the signal to background ratio must be of at least 5\%, everything evaluated separately in each energy bin.
As clearly visible in Fig.~\ref{fig:sensitivity_mono}, the sensitivity is significantly better for additional $\gamma$/h separation variables.
The improvement in sensitivity at $\approx$10\,TeV corresponds to the energy above which the \Palpha\ and $LCm$ calculations are available.

The $N_{\mu}$ estimates illustrate the potential of future $\gamma$/h separation methods at low energies
under an optimistic scenario, while the performance of~$LCm$ and \Palpha\ shows their impact at high energies.
The feasibility of $N_{\mu}$ identification is investigated, e.g., in Ref.~\citep{Conceicao:2021xgn}; however,
interestingly, the $LCm$~variable even surpasses $N_{\mu}$ at ultra-high energies as discussed in Ref.~\citep{LCmPeV}.

\subsection{Stereoscopic performance}
\label{sec:stereo}

\begin{figure}
    \centering
    \includegraphics[width=1.0\linewidth]{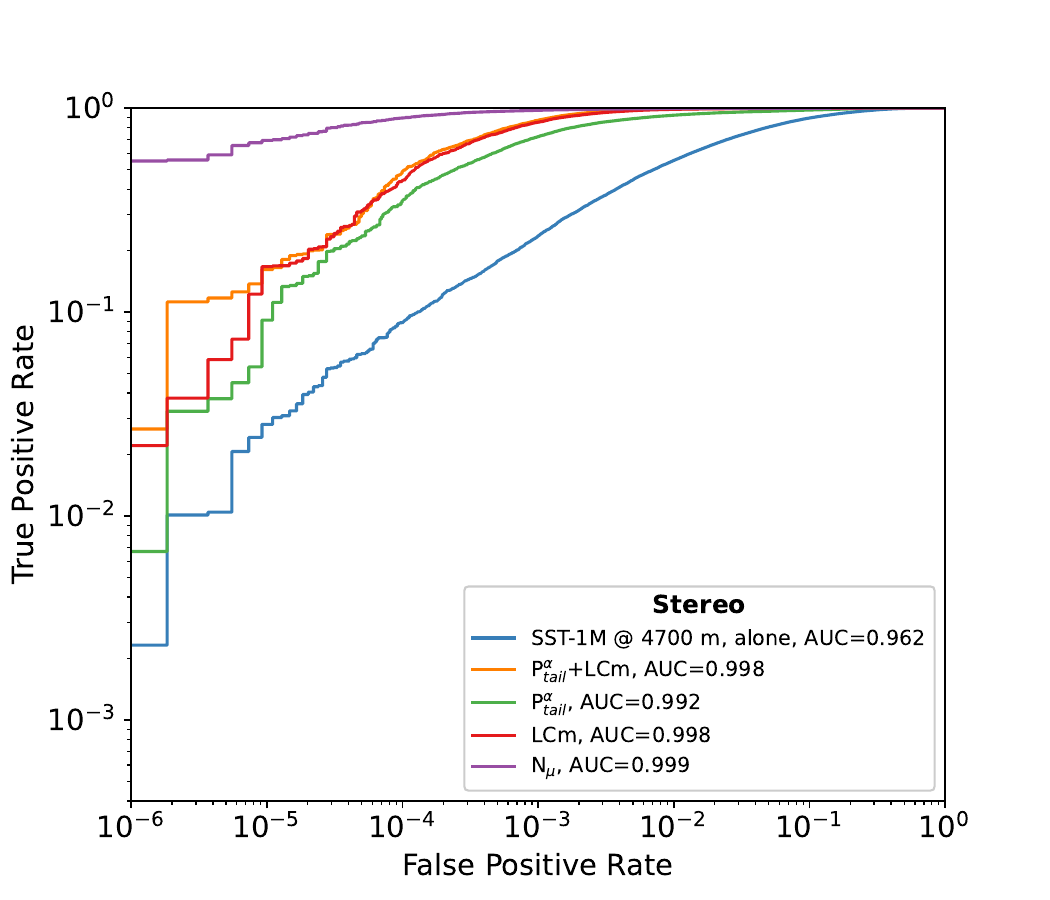}
    \caption{ROC curves in the stereoscopic observation regime.
    Shown for a pair of SST-1M telescopes alone (blue) and for SST-1Ms together with $\gamma$/h discrimination parameters similarly to Fig.~\ref{fig:roc_mono}.  
    All curves are shown for energies above 10\,TeV.}
    \label{fig:roc_stereo}
\end{figure}

\begin{figure}
    \centering
    \includegraphics[width=1.0\linewidth]{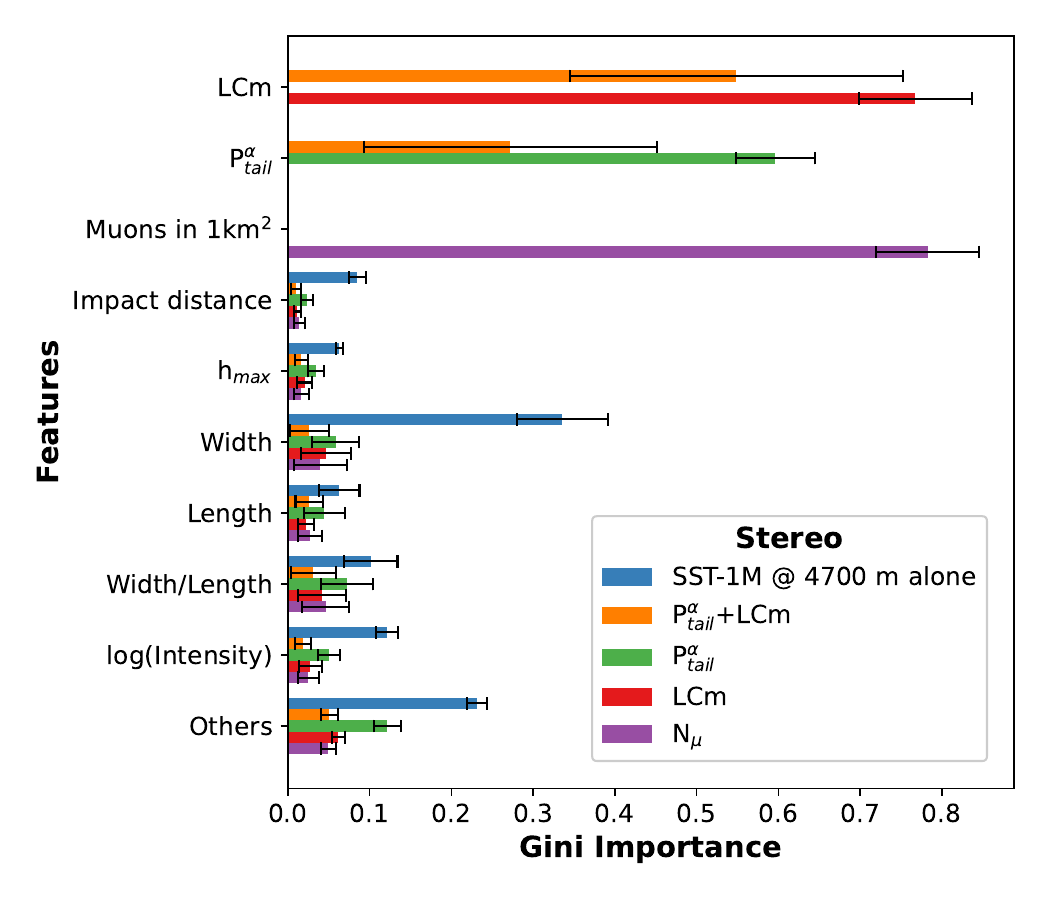}
    \caption{Gini importance of features used in the Random Forests classifiers when stereoscopic observation is assumed.
    Identical color coding as in Fig.~\ref{fig:gini_mono} is used, as well as the summed quantities in Others are the same.
    For simplicity, we show the output for the northern telescope only, the results of the southern one are nearly identical.}
    \label{fig:gini_stereo}
\end{figure}

\begin{figure}
    \centering
    \includegraphics[width=1.0\linewidth]{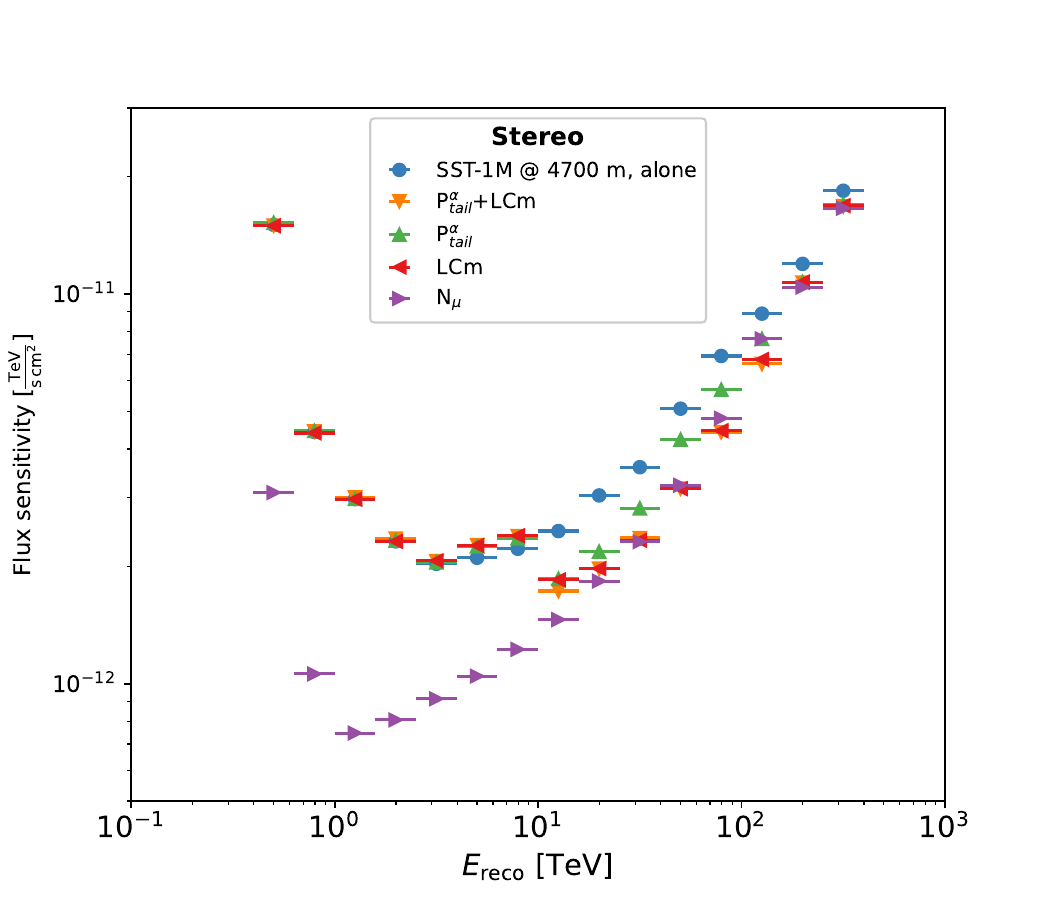}
    \caption{Flux sensitivity for two SST-1M telescopes (blue) and SST-1M telescopes with the additional information from the particle-detector array.
    The color coding and the observation parameters are the same as in Fig~\ref{fig:sensitivity_mono}.}
    \label{fig:sensitivity_stereo}
\end{figure}

In the case of two SST-1M telescopes, additional parameters derived from the stereoscopic analysis, the impact distance and $h_{\rm{max}}$, were included in the RF reconstruction.
Similarly to the monocular case, ROC curves, Gini importance, and sensitivities for the stereoscopic case are shown in Figs.~\ref{fig:roc_stereo}--\ref{fig:sensitivity_stereo}, respectively.
The improvement in sensitivity resulting from enhanced $\gamma$/h discrimination is less pronounced than that observed in the monocular case, primarily due to the inherently more precise reconstruction of the standalone SST-1M stereoscopic configuration, which leaves less room for improvement.
Nevertheless, similar to the monocular performance, $LCm$ surpasses $N_{\mu}$ at energies around 100\,TeV.

\begin{figure}
    \centering
    \includegraphics[width=1.0\linewidth]{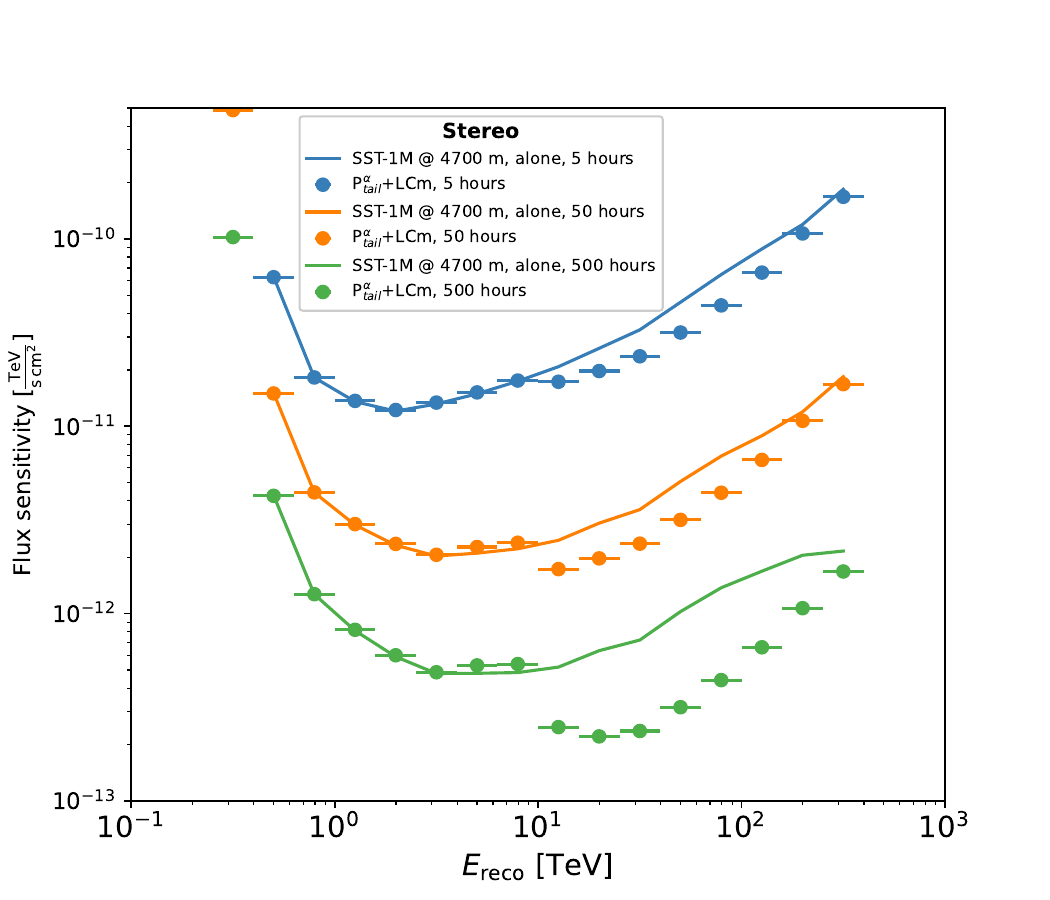}
    \caption{Evolution of the flux sensitivity with increasing observation time for the stereoscopic analysis enhanced with $LCm$+\Palpha\ $\gamma$/h separation.
    The same point-like source as in Figs.~\ref{fig:sensitivity_mono} and \ref{fig:sensitivity_stereo} was used.
    The lines represent sensitivities without WCD-array information.
    }
    \label{fig:sensitivity_evo_stereo}
\end{figure}

To illustrate the scaling of sensitivity with observation time, we evaluated the combined $LCm$+\Palpha\ sensitivity for 5, 50, and 500\,h in Fig.~\ref{fig:sensitivity_evo_stereo}.
In the ultra-high-energy regime (>100\,TeV), the measurement is effectively background-free, thus being limited by the requirement of 10 detected photons in each energy bin.

\section{Other benefits of hybrid detection}
\label{sec:hybrid}

Besides the improvement in $\gamma$/h separation of the SST-1M combined with information from the particle detectors, we stress additional scientific benefits of a hybrid combination of an array like SWGO and SST-1Ms.

\begin{figure}
    \centering
    \includegraphics[width=1.0\linewidth]{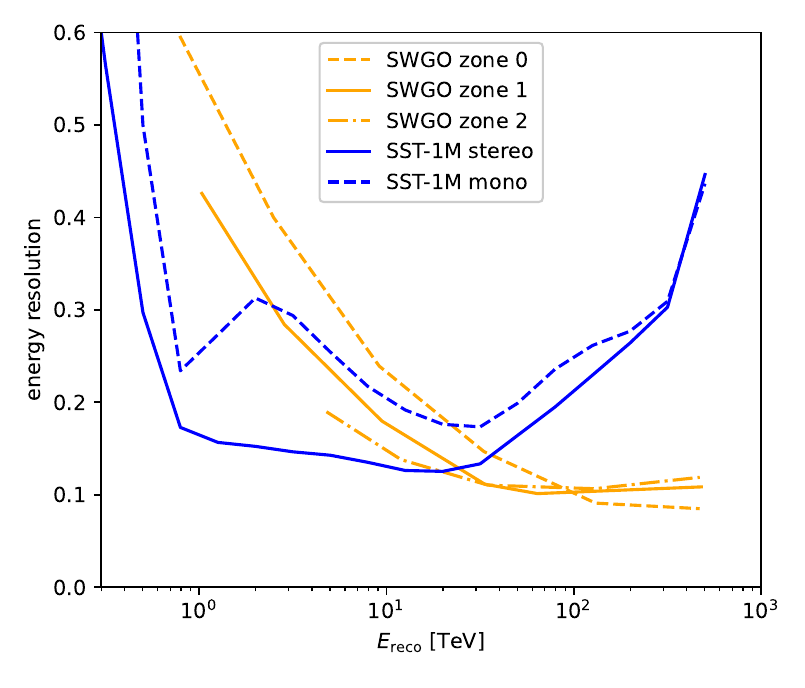}
    \caption{Resolution of the energy reconstruction of gamma showers, defined as the 68\% quantile of $|E_{\rm{reco}}-E_{\rm{true}}|/E_{\rm{true}}$.
    The estimates of individual SWGO zones (orange) come from Ref.~\citep{SWGO:2025taj}, while the SST-1M results (blue) are derived from simulations described in Section~\ref{sec:sim}.}
    \label{fig:energy_resolution}
\end{figure}

Up to energies of about 10\,TeV, the SST-1M telescopes operated at the SWGO site could provide an effective cross-calibration tool, offering better energy resolution, see Fig.~\ref{fig:energy_resolution}, and lower systematic uncertainty (about 10\% or below as estimated in Ref.~\citep{SST-1M}).
This could help to refine the absolute energy scale of the primary SWGO detector across the whole energy range.
In comparison, the main SWGO array is expected to achieve an energy resolution of about 15\% or worse, see Ref.~\citep{SWGO:2025taj}, and an energy-scale uncertainty of about 10\% or worse, which is strongly dependent on energy and observation time when inferred from the Moon-shadow method \citep{PhysRevD.104.062007}.
Around 10\,TeV, the most relevant part of SWGO is zone 1 due to its effective area.

\begin{figure}
    \centering
    \includegraphics[width=1.0\linewidth]{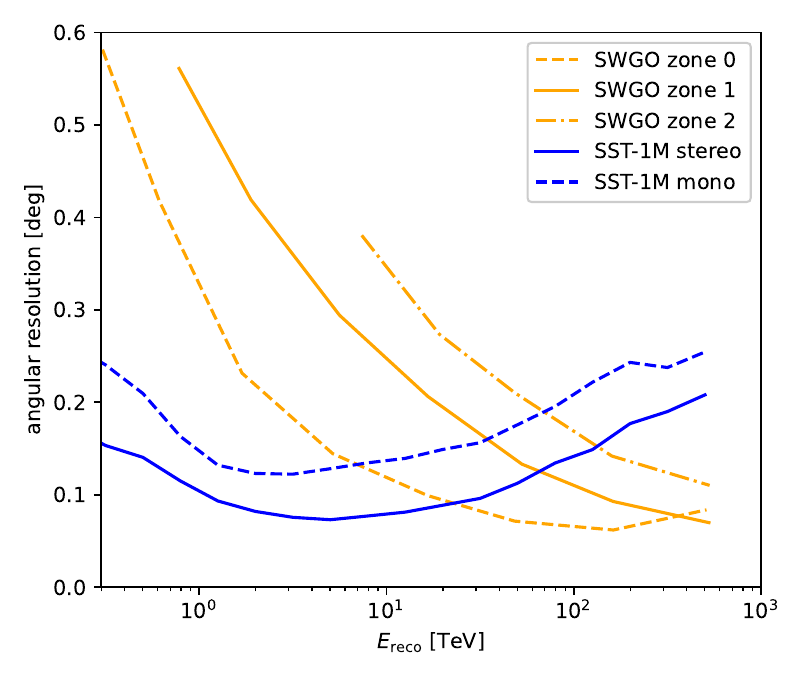}
    \caption{Angular resolution defined as the 68\% quantile of the angular distance between the reconstructed and the true direction of incident gammas.
    The color coding is the same as in Fig.~\ref{fig:energy_resolution}.}
    \label{fig:angular_resolution}
\end{figure}

The SST-1M data can constrain the air-shower geometry more than SWGO for energies $\approx$1--10\,TeV, where the SST-1M reconstruction resolutions are getting continuously better, see Figs.~\ref{fig:energy_resolution}~and~\ref{fig:angular_resolution} together with Ref.~\citep{SST1M_performance_altitudes_icrc2025}.
It may potentially improve the overall angular resolution for coincident events. The core resolution of the main SWGO array reaching values below 10~meters above 3\,TeV, see Ref.~\citep{SWGO:2025taj}, together with the angular resolution of the SST-1M telescopes shown in Fig.~\ref{fig:angular_resolution}, could provide significant improvement in the air-shower geometry reconstruction for the combined analysis. Another important aspect would also be an independent cross-check of the angular resolution of the SWGO array using the SST-1M telescopes.

Furthermore, the hybrid setup enables immediate follow-up observations by SST-1M on transient or variable sources detected by the wide-field SWGO, providing crucial spectral and morphological data due to its fine angular resolution shown in Fig.~\ref{fig:angular_resolution}, complementing the SWGO survey capabilities presented in Ref.~\citep{SWGO:2025taj}.
Although follow-up observations can be made from other sites as well, the integration of IACTs into SWGO can decrease the reaction time to a few tens of seconds.

A further consideration is the duty cycle: while WCD arrays can operate with near-continuous uptime, IACT observations are traditionally restricted to clear nights and low-moon conditions, resulting in a significantly lower effective duty cycle.
Consequently, performance gains enabled by IACT information are available only during a fraction of the total observation time.
Nevertheless, modern IACT cameras employing silicon photomultipliers (SiPMs), see Ref.~\citep{2017EPJC...77...47H}, can tolerate high night-sky background levels and may extend operations into bright-moon conditions, partially mitigating this limitation.
As the camera of SST-1M is based on SiPMs, it would be possible to test operation conditions under high moonlight as the additional information from the particle array may allow to reduce the degradation in performance.

\section{Technical aspects of operating SST-1M telescopes at the SWGO site}
\label{sec:tech}
To ensure mutual compatibility between SST-1M and SWGO operation, specific requirements need to be met.
A fenced area of approximately $20 \times 30\,\mathrm{m}^2$ is required around the telescope structure to restrict unauthorized access during the operation.
Furthermore, to access all potential sources within the operational limits from 0\textdegree~to 50\textdegree~in zenith and 0\textdegree~to 360\textdegree~in azimuth in a clear, free field-of-view, the entire volume must be free from obscuration by any nearby infrastructure.
These requirements effectively prevent placing the telescopes inside the highest-density part of the array.

According to the measurements performed at the Pampa La Bola site in 4700\,m altitude and analyzed during the SWGO site selection process, the telescope must be able to withstand long periods with temperatures between $-5\text{ \textdegree C}$ and $0\text{ \textdegree C}$ with short-term extremes dropping below $-20\text{ \textdegree C}$ during winter.
The reduced air pressure at high altitude diminishes convective heat transfer efficiency, therefore influencing the cooling performance required for the camera system during operation.
Furthermore, drive motors must operate reliably at low temperatures.
The current SST-1M setup is not designed for so low air pressure and temperature, but the development of the necessary adaptation is ongoing, including the refurbishment of the cooling system and potential use of heaters for the motors.

A high wind speed on the site can, in addition to operating restrictions, also influence the durability of the mirror segments.
The use of back-coated mirror technologies introduced in Ref.~\citep{back_coated} will mitigate substantially the progressive loss of the mirror reflectivity that is caused by the impact of small dust particles carried by high-velocity winds, despite it will reduce slightly the overall optical efficiency.

An adequate source of electricity must be present on site.
The IT system requires a constant power consumption of 1\,kW, which is independent of the number of telescopes.
For a single SST-1M telescope, the power demand is estimated at 2\,kW in standby mode and approximately 4\,kW during operation, with a peak instantaneous power level of up to 8.5\,kW.
Assuming a daily duty cycle of 14 hours in standby and 10 hours in operation, the resulting average daily energy consumption is approximately 70\,kWh.

Another critical operational aspect is the required data transfer rate and storage capacity.
A single SST-1M telescope stores approximately 500 GB per observation night, whereas the daily data volume recorded by a WCD array is about 2.5 TB according to Ref.~\citep{LHAASO:2024kbg}.
Given that IACTs cannot operate every night, they account for an additional data-handling capacity of approximately 10\% per telescope.

\section{Conclusions}
\label{sec:sum}

In this work, we estimated the benefit of locating the SST-1M telescopes in the center of a particle-detector array. We evaluate this through different $\gamma$/h discriminating variables and stereo and monocular mode.

The monocular and stereoscopic flux sensitivities of SST-1M have been improved by about 60\% and 30\% above 10\,TeV, respectively, due to the enhanced $\gamma$/h separation capability when the $LCm$ parameter of the WCD array was used.

The hybrid configuration allows for cross-calibration of reconstructed variables and independent IACT measurements can confirm the outputs of Deep-Learning $\gamma$/h algorithms \citep{Capistran:2025zql} that will be used in the new generation of WCD-based observatories.
Moreover, classification of charged cosmic-ray primaries can be studied.
However, investigation of these further possibilities is beyond the scope of the present study, in which we focused on the improvement of SST-1M performance allowed by more precise $\gamma$/h separation provided by the particle-detector array.
The comparison with SWGO was presented only at the level of angular and energy resolutions, while we stressed benefits of mutual cooperation.

Because no direct interference of SST-1M with the SWGO operation is expected, drawbacks of placing SST-1M at the SWGO site are mainly of technical origin.
The main technical aspects that need to be considered for such a high-altitude observatory equipped with the SST-1M telescope(s) are the new design for cooling of the camera, the drive operation, wind-speed alerts to park the telescope in time, and an extra energy consumption of about 70 kWh a day.
The atmospheric conditions on the site would also need to be monitored to allow for efficient SST-1M measurements.
In particular, the aerosol content and cloud presence in the FoV of the telescopes must be known, for which the FRAM telescope \citep{Aab_2021,Ebr_2021} can be used.

\section*{Acknowledgments}

This publication was created as part of the projects funded in Poland by the Minister of Science based on agreements number 2024/WK/03 and DIR/\-WK/2017/12. The construction, calibration, software control and support for operation of the SST-1M cameras is supported by SNF (grants CRSII2\_141877, 20FL21\_154221, CRSII2\_160830, \_166913, 200021-231799), by the Boninchi Foundation and by the Université de Genève, Faculté de Sciences, Département de Physique Nucléaire et Corpusculaire. The Czech partner institutions acknowledge support of the infrastructure and research projects by Ministry of Education, Youth and Sports of the Czech Republic (MEYS) and the European Union funds (EU), MEYS LM2023047, EU/MEYS CZ.02.01.01/00/22\_008/0004632, CZ.02.01.01/00/22\_010/0008598, Co-funded by the European Union (Physics for Future – Grant Agreement No. 101081515), and Czech Science Foundation, GACR 23-05827S.
The Portuguese contribution was supported by FCT - Funda\c{c}\~{a}o para a Ci\^encia e a Tecnologia, I.P. OE., by PRT/BD/154192/2022 [\href{https://doi.org/10.54499/PRT/BD/154192/2022}{DOI}] and 2023.18160.ICDT [\href{https://doi.org/10.54499/2023.18160.ICDT}{DOI}].

\bibliographystyle{elsarticle-num}
\bibliography{biblio.bib}

\end{document}